\documentclass[sigconf]{acmart}

\AtBeginDocument{%
  }

\usepackage{amsmath,amssymb,amsfonts}
\usepackage{algorithmic}
\usepackage{graphicx}
\usepackage{textcomp}
\usepackage{xcolor}
\usepackage{url} 
\usepackage{multirow}
\usepackage{booktabs}
\usepackage{tcolorbox}
\usepackage{diagbox}
\usepackage{tabularx}
\usepackage{microtype}
\usepackage{stackengine}  
\usepackage{listings}
\usepackage{subcaption} 
\usepackage{stackengine}
\usepackage{amsmath}
\usepackage{algorithm}
\usepackage{algorithmic}
\usepackage{wrapfig}
\usepackage[table]{xcolor} 
\usepackage{colortbl}

\newcommand{\findingboxx}[1]{
\begin{center}
\begin{tcolorbox}[colback=gray!11,
                  colframe=black,
                  boxrule=0.2mm,
                  width=0.45\textwidth,
                  arc=.8mm, auto outer arc,
                 ]
 #1
\end{tcolorbox}
\end{center}}

\setcopyright{acmlicensed}
\copyrightyear{2025}
\acmYear{2025}
\acmDOI{XXXXXXX.XXXXXXX}

\begin{document}

\title{OFP-Repair: Repairing Floating-point Errors via Original-Precision Arithmetic}

\author{Youshuai Tan}

\affiliation{%
  \institution{The Hong Kong University of Science and Technology (Guangzhou)}
    \country{China}
}
\email{ytan387@connect.hkust-gz.edu.cn}

\author{Zishuo Ding}
\authornote{Corresponding author.}
\affiliation{%
  \institution{The Hong Kong University of Science and Technology (Guangzhou)}
  \country{China}
  }
\email{zishuoding@hkust-gz.edu.cn}

\author{Jinfu Chen}
\affiliation{%
  \institution{Wuhan University}
  \country{China}}
  \email{jinfuchen@whu.edu.cn}

\author{Weiyi Shang}
\affiliation{%
 \institution{University of Waterloo}
 \country{Canada}}
 \email{wshang@uwaterloo.ca}

\renewcommand{\shortauthors}{Trovato et al.}

\begin{abstract}
Errors in floating-point programs can lead to severe consequences, particularly in critical domains such as military, aerospace, and financial systems, making their repair a crucial research problem. In practice, some errors can be fixed using original-precision arithmetic, while others require high-precision computation. Developers often avoid addressing the latter due to excessive computational resources required. However, they sometimes struggle to distinguish between these two types of errors, and existing repair tools fail to assist in this differentiation. Most current repair tools rely on high-precision implementations, which are time-consuming to develop and demand specialized expertise. Although a few tools do not require high-precision programs, they can only fix a limited subset of errors or produce suboptimal results.

To address these challenges, we propose a novel method, named OFP-Repair. Our method identifies original-precision-repairable errors by computing the condition number of the program with respect to its inputs. We then construct a unified repair framework using series expansion. ACESO is the only existing tool that neither requires nor relies on high-precision programs for both detection and repair, making it our primary baseline. On ACESO’s dataset, our patches achieve improvements of three, seven, three, and eight orders of magnitude across four accuracy metrics. In real-world cases, our method successfully detects all five original-precision-repairable errors and fixes three, whereas ACESO only repairs one. Notably, these results are based on verified data and do not fully capture the potential of OFP-Repair. To further validate our method, we deploy it on a decade-old open bug report from GNU Scientific Library (GSL), successfully repairing five out of 15 bugs. The developers have expressed interest in our method and are considering integrating our tool into their development workflow. We are currently working on applying our patches to GSL. The results are highly encouraging, demonstrating the practical applicability of our technique.

\end{abstract}




\setcopyright{none} 
\settopmatter{printacmref=false} 
\maketitle

\section{Introduction}
\label{Introduction}
Floating-point programs are the basis of modern science and engineering and their errors can lead to serious consequences, including military applications~\cite{skeel1992roundoff}, aerospace engineering~\cite{lions1996flight}, and financial systems~\cite{weisstein1999roundoff, quinn1983ever}. Therefore, many works have been conducted to detect the potential floating-point errors~\cite{barr2013automatic, ma2022numfuzz, zou2019detecting, yi2019efficient, zhang2024hierarchical, yi2024fpcc}. Upon error detection, subsequent repair proves equally critical.

In practical scenarios, developers address numerical accuracy errors through two distinct strategies~\cite{di2017comprehensive}: 1) utilizing high-precision floating-point arithmetic; and 2) restructuring numerical expressions while maintaining the original computational precision. Developers generally avoid increasing numerical precision as a corrective measure, as this approach necessitates substantial computational resources and significantly prolongs execution time~\cite{larsson2013exploring}. For example, NumPy issue \#1063 was closed without applying the fix because higher precision is too expensive~\cite{di2017comprehensive}. In contrast, developers fix the SciPy issue \#3545 by utilizing different but mathematically equal expressions. However, although developers prefer the cases that can be repaired using original-precision arithmetic, no currently available instrumentation facilitates developers' assessment of whether errors are repairable through original-precision numerical computations alone. For example, the developers initially attempted to resolve SciPy issue \#6368 by modifying computational expressions, but subsequent results revealed that the bug could only be fixed with higher-precision arithmetic.

Although prior research has proposed many methods for repairing floating-point errors, they lack the capability to effectively identify and resolve the errors that can be repaired using only original-precision arithmetic: \textbf{1) Limitation 1: Both the detection and repair processes necessitate the use of high-precision programs.} Transforming original programs into high-precision versions needs profound knowledge of math and numerical analysis~\cite{wang2016detecting}. \textbf{2) Limitation 2: Absence of a unified repair paradigm for original-precision-repairable errors.} Although some tools~\cite{panchekha2015automatically, damouche2017salsa, xu2024arfa, wang2019global} utilize mathematically identical transformations to repair errors, they are only capable of handling some of the original-precision-repairable errors because they rely heavily on templates. \textbf{3) Limitation 3: Lack of diagnostic capability for such errors.} This pain point manifests concretely in practical development scenarios, as demonstrated by SciPy issue \#6368.

To address these limitations, we propose OFP-Repair (\textbf{Repair} \textbf{F}loating-\textbf{P}oint errors by using \textbf{O}riginal-precision arithmetic). We employ finite difference methods to compute numerical derivatives and derive the condition numbers of the target programs, thereby assessing whether the bug can be repaired by using original-precision computations. Subsequently, our method uses Taylor series to construct patches. 

Since ACESO is currently the only approach that does not employ high-precision program repair, we evaluate our method's performance by comparing it with ACESO on the latter's dataset. Experimental results demonstrate that our method can successfully identify all the original-precision-repairable errors. Moreover, our patches achieve significantly higher precision than those of ACESO across all evaluated metrics. Specifically, the average maximum absolute error of all functions on the stable area is \(4.11\times 10^{-16}\) (vs. ACESO's \(2.45 \times 10^{-13}\)), demonstrating a three-order-of-magnitude improvement. For maximum relative error on the stable area, we attain \(7.47\times 10^{-16}\) compared to ACESO's \(2.74 \times 10^{-9}\), representing a seven-fold enhancement in precision. Similarly, in the decayed area, our maximum absolute error of \(2.13\times 10^{-16}\) outperforms ACESO's \(2.45\times 10^{-13}\) by three orders of magnitude. Our maximum relative error of \(3.73\times 10^{-15}\) in the decayed region shows an eight-order-of-magnitude improvement over ACESO's \(5.74\times 10^{-07}\). We further evaluate our method on five real-world original-precision-repairable errors collected by Franco et al.~\cite{di2017comprehensive}. We can identify all five bugs and successfully repair three of them. In contrast, ACESO is only able to fix one bug.

In addition to applying our method to previously validated original-precision-repairable errors, we also attempt to repair open bug reports. We believe this evaluation strategy better demonstrates the effectiveness of our method, as successfully repairing such bugs suggests that our method is highly effective and has the potential for practical adoption in future software development. We use our method to repair bugs from an open GNU Scientific Library (GSL) bug report that was posted over a decade ago~\footnote{\protect\url{http://savannah.gnu.org/bugs/?43259}}. OFP-Repair is able to reduce the numerical errors in five out of the 15 bugs reported in this bug report. For three of these bugs, we successfully reduce the relative error to below \(1 \times 10^{-16}\). For the other two bugs, we reduce the relative error by one and 15 orders of magnitude, respectively. These results indicate that, if our method had been available to developers at the time, the bugs could likely have been resolved instead of remaining open for more than ten years. The GSL developers have shown interest in adopting our method.

The replication package, including the detection process and our patches, is available at the following link~\footnote{\protect\url{https://github.com/abs-hello/fixing}}. In summary, our work makes three main contributions:
\begin{itemize}
\item We introduce OFP-Repair, a novel method designed to effectively detect original-precision-repairable errors and repair them.
\item We evaluate OFP-Repair on a diverse set of benchmarks, including datasets of ACESO and real-world bugs from Franco et al.~\cite{di2017comprehensive}.
\item We successfully reduce relative errors of five bugs from a bug report that has remained open for over a decade.
\end{itemize}

To ensure clarity of key concepts utilized in this manuscript, we present the following definitions: In this context, we define the \textbf{original-precision-repairable errors} as the floating-point errors that can be fixed by using original-precision floating-point representations. 

\textbf{Work organization.} The remainder of this work is structured as follows. Section~\ref{background} provides the necessary background for our research. In Section~\ref{Motivation}, we introduce the motivation of our work and discuss the shortcomings of current approaches. Section~\ref{OFP-Repair} details the design and implementation of OFP-Repair. The experimental evaluation is presented in Section~\ref{eva}. We present the results of our method when applied to the bugs in the bug report in Section~\ref{case_study}. We analyze potential limitations in Section~\ref{threat} and review related work in Section~\ref{related-work}. Finally, Section~\ref{conclusion} summarizes our contributions and concludes the work.

\section{Background}
\label{background}

In this section, we provide an overview of the background related to Taylor series, floating-point representation, and cancellation.

\subsection{Taylor series}
\label{taylor_series}
\begin{figure}[htbp]
    \centering
    \includegraphics[width=0.48\textwidth]{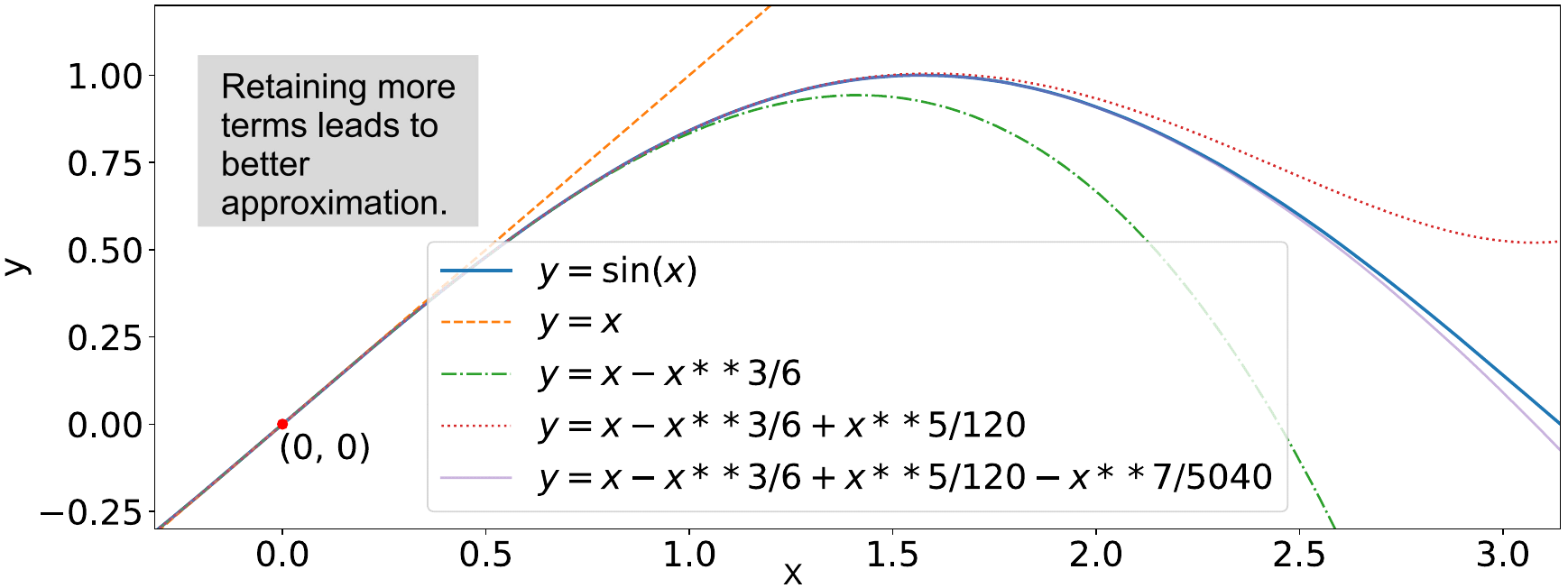}
    \vspace{-0.5cm}
    \caption{The approximation performance of Taylor series.}
    \vspace{-0.5cm}
    \label{sin}
\end{figure}

Taylor series are widely used in science and engineering~\cite{stewart2012calculus, foy1976position}. Equation~\ref{taylor} represents the Taylor series of the function \(f\) at \(a\). The Taylor series can be employed to approximate functions in the vicinity of the point \(a\) based on the initial several terms, exhibiting a rapid rate of convergence~\cite{stewart2012calculus}. Programmers prefer using polynomials to approximate the values of functions, as polynomials are among the simplest types of functions. For example, if we calculate the value of \(\sin(x)\) for \(x\) near 0, we have: \(\sin(x) = \sin(0) + \frac{\cos(0)}{1!} (x - 0) + \frac{-\sin(0)}{2!} (x - 0)^2 + \frac{-\cos(0)}{3!} (x - 0)^3 + \frac{\sin(0)}{4!} (x - 0)^4 + \frac{\cos(0)}{5!} (x - 0)^5 \cdots .\) Since we can directly obtain the values of \(\sin(0)\) and \(\cos(0)\), the equation becomes: \(\sin(x) = x - \frac{x^3}{3!} + \frac{x^5}{5!} + \cdots .\) Figure~\ref{sin} shows that retaining more terms in the Taylor series leads to better approximation performance.

\begin{equation}
\begin{aligned}
f(x) 
&= \sum_{n=0}^\infty \frac{f^{(n)}(a)}{n!} (x - a)^n\\
&= f(a) + \frac{f'(a)}{1!} (x - a) + \frac{f''(a)}{2!} (x - a)^2 + \cdots
\label{taylor}
\end{aligned}
\end{equation}

 The use of Taylor series for function approximation is a common approach in numerical software. In the GNU Scientific Library~\footnote{\protect\url{https://www.gnu.org/software/gsl/doc/html/specfunc.html\#trigonometric-functions}}, the \(gsl\_sf\_sin\_e\) function returns results computed using its Taylor series for inputs with an absolute value less than 1.2207031250000000e-04. Since only the first two terms of the Taylor series are retained, the applicable domain of the Taylor approximation in \(gsl\_sf\_sin\_e\) is limited.
 

\subsection{Floating-point representation}
Floating-point arithmetic had become widely adopted as a standard computational method by the mid-1950s~\cite{overton2001numerical}. To achieve a unified floating-point system, a standard for binary floating-point representation and arithmetic was established through the cooperation between academic computer scientists and microprocessor chip designers. The standard was published in 1985, which was known as IEEE (Institute for Electrical and Electronics Engineers) 754.

The IEEE floating-point numbers can be represented in a standardized format consisting of three key components:
\begin{equation}\pm(b_0.b_1b_2\ldots b_{p-1})_2\times2^E\text{, where }
\end{equation}
\begin{itemize}
    \item The sign \(\pm\) indicates whether the number is positive or negative;
    \item The binary fraction \(b_0.b_1b_2\ldots b_{p-1}\) is the significand (or mantissa) part, with precision \(p\). In the case of normalized numbers, the leading bit \(b_0\) is implicitly 1;
    \item The exponent \(E\) is computed by \(E=e-bias\), where \(e\) is the biased \(m\)-bit exponent value, and the \(bias\) is given by \(bias=2^{m-1}-1\). 
\end{itemize}

The IEEE standard defines two basic representation formats, \textit{single} and \textit{double}. Table~\ref{format} illustrates the format of the three parts. The standard also advocates the inclusion of an extended precision format, characterized by a minimum of 15 bits allocated to the exponent and at least 63 bits dedicated to the significand.

\begin{table}[tb]
\centering
\footnotesize
\caption{The bit allocation of IEEE standard floating-point numbers.}
\vspace{-0.3cm}
\label{format}
\begin{tabular}{lccr}
\toprule
 &  \textbf{Sign} & \textbf{Exponent}&\textbf{Significand}\\
\midrule
 \(Single\) precision (32 bits) & 1  & 8 & 23 \\
 \(Double\) precision (64 bits) &  1& 11 & 52 \\
\bottomrule
\end{tabular}
\vspace{-0.5cm}
\end{table}

\subsection{Cancellation}
\label{back_cancellation}
\begin{figure}[htbp]
    \centering
    \vspace{-0.4cm}
    \includegraphics[width=0.48\textwidth]{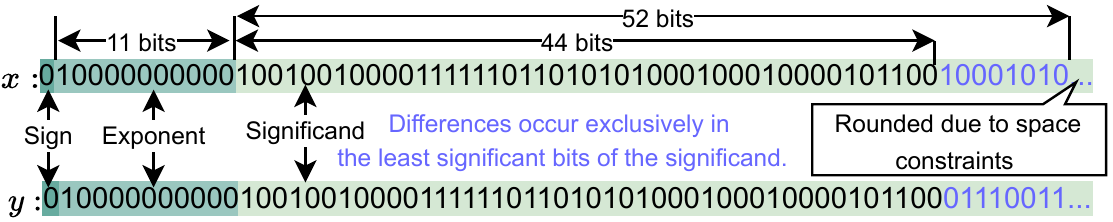}
    \vspace{-0.5cm}
    \caption{The double-precision floating-point representations of \(x\) and \(y\).}
    \vspace{-0.7cm}
    \label{cancellation}
\end{figure}

Cancellation, which arises when performing subtraction between two nearly identical floating-point numbers~\cite{overton2001numerical}, is one kind of operation which can result in significant errors for floating-point programs~\cite{zou2019detecting}. Analogously, it can also occur during the addition of two numbers with opposing signs but approximately equal magnitudes. For example, the difference between the two real numbers, \(x = 3.1415926535897\textbf{3}\) and \(y = 3.1415926535897\textbf{2}\), is \(1 \times 10^{-14}\). However, if we conduct the subtraction operation using double-precision floating-point numbers, the result is \(1.021405182655144 \times 10^{-14}\), leading to a relative error of about 0.0214, which is significant in many scientific applications. The observed accuracy degradation can be primarily ascribed to the constrained bit allocation for the significand component. As evidenced in Figure~\ref{cancellation}, a substantial cancellation of 44 significand bits occurs, resulting in merely eight effective precision bits remaining. In such a case, the computed result contains only eight correct bits in the significand component, whereas a double-precision floating-point number retains 52 significand bits. Consequently, the relative error between the computed result and the oracle value becomes substantial.

\section{Motivation}
\label{Motivation}
The first systematic study of numerical bugs in real-world software systems was carried out by Franco et al.~\cite{di2017comprehensive}. This section first presents our research motivation emerging from this systematic study. Then we introduce an example to demonstrate that certain floating-point errors can be effectively repaired using original-precision floating-point numbers without resorting to high-precision floating-point numbers. Finally, we examine existing error-repairing methods, highlighting their limitations in repairing such original-precision-repairable errors. 

Repairing floating-point errors without resorting to high-precision arithmetic represents a significant advancement, as high-precision programs developments typically involve substantial work~\cite{wang2016detecting} and high-precision programs tend to cost prolonged time~\cite{larsson2013exploring}. Furthermore, certain computational environments may not support high-precision floating-point representations.

\subsection{Why choose our method? Insights from a real-world numerical bugs study}
\label{why_our_method}
Franco et al.~\cite{di2017comprehensive} have conducted the first systematic investigation of numerical bugs encountered in real-world software systems, including NumPy\footnote{~\url{http://www.numpy.org/}}, SciPy\footnote{~\url{http://www.scipy.org/}}, LAPACK\footnote{~\url{http://www.netlib.org/lapack/}}, GNU Scientific Library\footnote{~\url{https://www.gnu.org/software/gsl/}}, and Elemental~\cite{poulson2013elemental}. Through rigorous analysis, they identified 269 numerical bugs and developed a novel taxonomy for classifying these bugs. One significant category is accuracy bugs due to rounding or truncation errors. Their detrimental effects are substantial, as they evade detection by compilers (e.g., underflow conditions), potentially leading users to obtain erroneous results. Based on their exploratory findings, developers address the accuracy errors through two primary approaches: one involves repairing the bug while maintaining the original precision; the other requires increasing the floating-point precision within the program. 

Developers often neglect accuracy errors requiring precision enhancement, as only a minimal subset of inputs triggers these errors~\cite{zou2019detecting}, and precision augmentation significantly degrades computational efficiency. For instance, NumPy issue \#1063 was closed without fixing. Interestingly, developers can face uncertainty in determining whether an error can be resolved within original precision constraints. The SciPy issue \#6368 exemplifies this phenomenon: developers initially attempted to address the error by modifying the computation order, but ultimately determined that extended-precision floating-point arithmetic was necessary to achieve correct results. Furthermore, developers tend to employ the Remez algorithm\footnote{~\url{https://github.com/scipy/scipy/issues/4034}}, which requires high-precision programs to obtain oracles and conduct approximations, to perform a repair. Therefore, we aim to design a method capable of identifying and repairing original-precision-repairable errors without relying on high-precision floating-point arithmetic.

\subsection{Original-precision-repairable errors}

Certain instances of errors can be significantly mitigated through the application of series expansion without using high-precision arithmetic. Consider the function \(\sin(x + \epsilon) - \sin(x)\), where the values of \(x\) and \(\epsilon\) are 2.13 and \(1 \times 10^{-6}\), respectively. We evaluate this function using the C++ Mathematical Library with double-precision arithmetic. For high-precision computation implementation, we utilize the \textit{mpmath}~\cite{mpmath}. Since the values of \(\sin(x + \epsilon)\) and \(\sin(x)\) are close to each other, the relative error value is \(1.1095 \times 10^{-10}\), demonstrating the cancellation error. According to the Taylor series (cf. Section~\ref{background}), we take \(\sin(x + \epsilon)\) as the point of \(\sin\) near \(x\) and have: \(\sin(x+\epsilon) = \sin(x) + \frac{\cos(x)}{1!} (\epsilon) + \frac{-\sin(x)}{2!} (\epsilon)^2 + \frac{-\cos(x)}{3!} (\epsilon)^3 + \frac{\sin(x)}{4!} (\epsilon)^4 + \frac{\cos(x)}{5!} (\epsilon)^5 \cdots .\) Interestingly, by omitting the \(\sin(x)\) term and calculating the remaining terms of the Taylor series directly, we can avoid the cancellation and mitigate the floating-point error of the target function. This approach achieves a relative error of \(1.6176 \times 10^{-16}\), illustrating a six-order-of-magnitude improvement in accuracy. We define this type of error as \textbf{original-precision-repairable errors}.

\subsection{Limitations of existing repair methods}

Floating-point errors are inevitable and can result in substantial inaccuracies with potentially harmful real-world consequences; therefore, many tools have been proposed to repair these errors. However, the existing tools cannot detect and repair the aforementioned repairable floating-point errors effectively. 

\begin{figure}[htbp]
    \centering
    \vspace{-0cm}
    \includegraphics[width=0.48\textwidth]{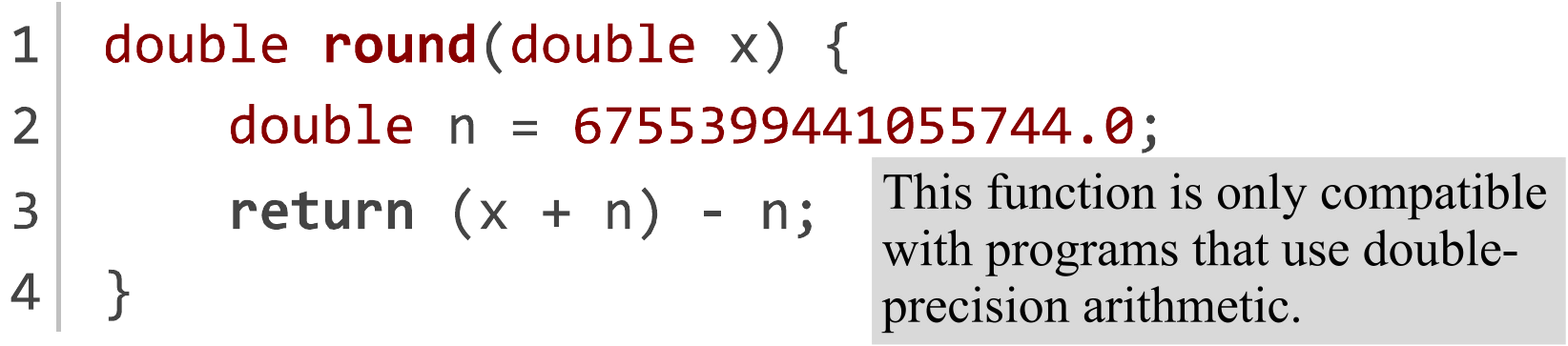}
    \vspace{-0.4cm}
    \caption{One precision-specific code snippet in GLIBC.}
    \vspace{-0.2cm}
    \label{motivation-example-one}
\end{figure}

\noindent \textbf{Limitation 1: Inability to repair floating-point errors without transitioning entire floating-point programs to high-precision versions.} Based on the observation that large floating-point errors tend to be triggered by inputs localized in small input intervals, AutoRNP~\cite{yi2019efficient} selects some points and corresponding oracles in the intervals as samples and utilizes linear approximation to construct patches. To obtain oracles of the sample points, the authors transform the complete program into a high-precision implementation. However, the transform processes require much expertise and effort, because directly increasing numerical precision (e.g., replacing double-precision (64-bit) floating-point numbers with 128-bit extended-precision format) can be ineffective and even introduce substantial errors~\cite{wang2016detecting}. For example, Figure~\ref{motivation-example-one} illustrates a simplified code piece in \(\exp\) function from GLIBC, which aims to round \(x\) to an integer~\cite{wang2016detecting}. It is noted that this function only works for double-precision floating-point numbers. If we execute this function using higher-precision data types (such as long double), we typically will not observe the rounding effects seen with double-precision types. This is because long double variables possess more bits in significand. Consequently, the computation will generally return the exact value of \(x\), violating the original semantics. Therefore, converting the entire program into a correct high-precision version requires significant time and extensive domain knowledge. Moreover, the execution time of high-precision programs is prolonged. For instance, execution using quadruple precision (128 bits) is nearly two orders of magnitude slower than double precision (64 bits)~\cite{larsson2013exploring}.

\noindent \textbf{Limitation 2: Only capable of resolving a subset of the errors.} A line of research improves the accuracy of floating-point programs by using mathematically equivalent transformations~\cite{panchekha2015automatically, damouche2017salsa, xu2024arfa, wang2019global}. For example, the value of \(\sqrt{x+1} - \sqrt{x}\) is equal to \(1/(\sqrt{x+1}+\sqrt{x})\) by rationalizing the numerator: \(\sqrt{x+1} - \sqrt{x}= \frac{(\sqrt{x+1} - \sqrt{x})(\sqrt{x+1} + \sqrt{x})}{\sqrt{x+1} + \sqrt{x}} = \frac{(\sqrt{x+1})^2 - (\sqrt{x})^2}{\sqrt{x+1} + \sqrt{x}} = \frac{1}{\sqrt{x+1} + \sqrt{x}}.\) Consequently, this transformation can avoid the cancellation cases when \(x\) is approaching 0. As for our illustrative example, Herbie~\cite{panchekha2015automatically} converts the original express to \((\sin(5\times 10^{-7}) \times 2.0) \times \cos((-5\times 10^{-7} - x))\) by using trigonometric transformations. However, these approaches rely on predefined templates, limiting their ability to cover all possible repairable floating-point errors. For instance, Herbie only supports some basic mathematical functions, such as \textbf{exp}, \textbf{log}, and \textbf{sin}, but cannot handle more complex ones like those provided by the GNU Scientific Library (GSL)~\footnote{\url{https://www.gnu.org/software/gsl/}}.

ACESO~\cite{zou2022oracle} is the only existing tool that performs both detection and repairing of floating-point errors without relying on high-precision floating-point arithmetic. They detect the erroneous operations by calculating atomic condition numbers~\cite{zou2019detecting}. Similar to AutoRNP, ACESO utilizes polynomial approximation to obtain patches. However, instead of calculating oracles of sample points by using high-precision programs, ACESO computes oracles by capturing the \textit{micro-structure} characterization of floating-point errors. Specifically, it selects points near the erroneous input within a carefully determined radius and computes their average output, which approximates the oracle value. Despite its innovation, ACESO’s repairing performance exhibits limitations in certain scenarios. For example, in our illustrative case, ACESO’s patch achieves a relative error of magnitude \( 10^{-9}\), which significantly falls short of the accuracy delivered by Herbie \( 10^{-16}\) and the Taylor series-based approach \( 10^{-16}\).

\noindent \textbf{Limitation 3: Failure to identify original-precision-repairable errors.} Effective error repair fundamentally requires precise error detection and localization, a capability currently lacking in state-of-the-art tools. The most closely related approaches are template-based repair methods, which nevertheless exhibit limited coverage of such errors. Furthermore, they lack theoretical foundations to understand these reparable cases.

To address these limitations, we propose our method, named OFP-Repair, to detect and repair the original-precision-repairable errors systematically and effectively.

\section{OFP-Repair: detect and repair original-precision-repairable errors}
\label{OFP-Repair}

\begin{figure*}[htbp]
\centerline{\includegraphics[width=0.95\textwidth]{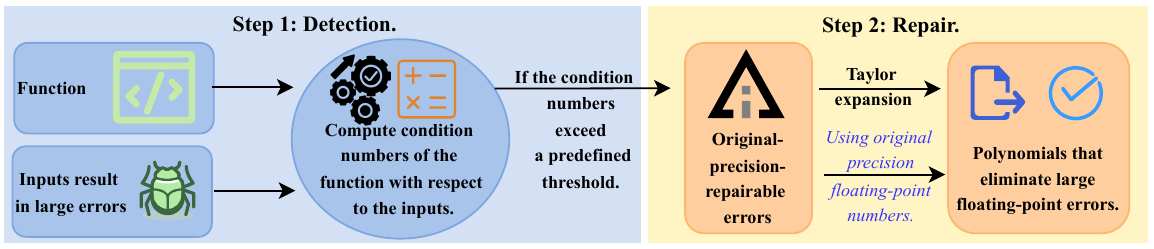}}
\vspace{-0.3cm}
\caption{An overview of OFP-Repair.}
\vspace{-0.2cm}
\label{framework}
\end{figure*}

In this section, we elaborate on our method, OFP-Repair. We begin by presenting the theoretical foundations of our method, elucidating the design rationale. Regarding error detection, we evaluate whether the condition number of the input with respect to the program is sufficiently small. Errors meeting this criterion are classified as original-precision-repairable errors. Subsequently, we derive corrective patches through series expansion. We use the illustrative function \(\sin(x + \epsilon) - \sin(x)\) in Section~\ref{Motivation} to show our detection and repair process. An overview of our method is shown in Figure~\ref{framework}.

\subsection{Foundations of OFP-Repair}
\label{foundations_of}

Before introducing the details of our method, this subsection establishes the theoretical foundations of our method. We first analyze the origins of large floating-point errors, demonstrating that they predominantly arise from cancellations (cf.~\autoref{back_cancellation}). Second, we show that programs can be converted into polynomial representations using Weierstrass approximation theorem~\cite{tao2016analysis}. 

\noindent \textbf{Significant errors stem from cancellations.} Before repairing errors, we investigate the root causes of floating-point inaccuracies. Zou et al.~\cite{zou2019detecting} proposed that substantial floating-point errors originate from atomic operations (such as \(+, \times, \sin, \cos\), and \(\exp\)) within large condition numbers. However, at the hardware/implementation level, these operations are ultimately composed of basic arithmetic operations including addition, subtraction, multiplication, and division. Notably, for these operations which may lead to significant condition numbers, only addition and subtraction are among the fundamental hardware-supported arithmetic operations~\cite{hassanien2009foundations}. When the condition numbers of addition and subtraction are large, cancellations appear in the program. Therefore, this analysis consequently demonstrates that the source of significant floating-point errors can be traced to cancellation effects. 

We take the \(\sin\) function of the GNU Scientific Library as an example to show that large floating-point errors of the atomic operations are attributed to cancellations. In~\autoref{taylor_series}, we introduce that the \(gsl\_sf\_sin\_e\) function simulates \(\sin\) by using Taylor series for inputs with small absolute values. Concerning large inputs, \(gsl\_sf\_sin\_e\) first reduces the input to the principal domain $[-\pi, \pi]$ and determines its octant location. When computing the remainder term, cancellations occur for inputs that are integer multiples of $\pi$ (excluding zero). Zou et al.~\cite{zou2019detecting} found the same domains that can lead to large condition numbers of \(\sin\) by using mathematical derivation. Therefore, cancellations lead to the large floating-point errors of \(\sin\) function.

\noindent \textbf{Programs are convertible to polynomial representations.} To begin with, we consider the following two well-known theorems:

\textbf{Theorem 1} (Arithmetic Operations Preserve Continuity Theorem~\cite{stewart2012calculus}): \textit{If \(f\) and \(g\) are continuous at \(a\) and \(c\) is a constant, then the following functions are also continuous at \(a\): \(f+g\), $f-g$, $cf$, $fg$, and $\frac{f}{g}$ $\text{if } g(a) \neq 0$.}

\textbf{Theorem 2} (Weierstrass Approximation Theorem~\cite{tao2016analysis}): \textit{If $[a, b]$ is an interval, $f \colon [a, b] \to \mathbb{R}$ is a continuous function, and $\varepsilon > 0$, then there exists a polynomial $P$ on $[a, b]$ such that $d_{\infty}(P, f) \leq \varepsilon$ (i.e., $|P(x) - f(x)| \leq \varepsilon$ for all $x \in [a, b]$).}

As previously discussed, arithmetic parts of floating-point programs can be fully decomposed at the low level into arithmetic operations (addition, subtraction, multiplication, and division) involving input variables and constants. Therefore, according to Theorem 1, any floating-point program is continuous within the domain of the corresponding branch. It is crucial to note that this continuity must be constrained within individual branches, as different branches may employ distinct expressions, thereby breaking continuity across branch boundaries. For example, \(gsl\_sf\_sin\_e\) employs distinct computational procedures for small and large absolute values. Furthermore, as established by Theorem 2, the original floating-point program can be equivalently represented as a polynomial within each branch domain. Since distinct polynomials can be constructed for different branches, we can conclude that floating-point programs are reducible to polynomial representations in general.

A critical observation is that while certain cancellation errors can be effectively mitigated through careful manipulation of original-precision arithmetic, others prove fundamentally irreducible within the precision constraint. Therefore, we assume that original-precision-repairable programs can be converted into polynomial representations that are free of cancellation.


\subsection{(Overcoming limitation 1 and 3) Step 1: Detection}
\label{overcoming_limitation_1and3}

In this subsection, we first present the design inspiration. Subsequently, we show the details to identify the original-precision-repairable errors.

\noindent \textbf{Design inspiration: How to isolate original-precision-repairable errors from general floating-point errors?} Prior to answering this question, we examine the mechanism by which ATOMU detects floating-point errors. In Section~\ref{back_cancellation}, we have demonstrated that the subtraction of two floating-point numbers with similar magnitudes results in significant errors from the floating-point arithmetic perspective. However, if the real numbers \(x\) and \(y\) in Figure~\ref{cancellation} are exactly representable as floating-point numbers (i.e., the truncated portions denoted by ellipses are all zeros), then the error becomes zero. Therefore, cancellation is highly probable, though not guaranteed, to produce substantial floating-point errors. To measure the magnitude of probability, ATOMU utilizes the condition number theory:
\vspace{-0.1cm}
\begin{equation}
\left|\frac{f(x+\Delta x)-f(x)}{f(x)}\right| \approx \left|\frac{\Delta x}{x}\right|\cdot\left|\frac{xf^{\prime}(x)}{f(x)}\right|
\end{equation}
where \(\Delta x\) denotes a small perturbation. \(\left|\frac{f(x+\Delta x)-f(x)}{f(x)}\right|\) and \(\left|\frac{\Delta x}{x}\right|\) mean relative errors of output and input, respectively. \(\left|\frac{xf^{\prime}(x)}{f(x)}\right|\) is the corresponding condition number, representing the error gain from input to output. For example, concerning our motivation example in Section~\ref{Motivation}, we take \(\sin(x + \epsilon)\) and \(\sin(x)\) as variables, respectively. The corresponding condition number with respect to \(\sin(x + \epsilon)\) is \(\left|\frac{\sin(x + \epsilon)}{\sin(x + \epsilon)-\sin(x)}\right|\), resulting in a value of \(9.0132 \times 10^{9}\). Due to the inherent inaccuracy of floating-point arithmetic, \(\sin(x + \epsilon)\) is likely to contain minor errors, which can result in significant error propagation in the final results. It should be noted that this amplification mechanism is mathematically derived---even if the subtraction operation itself is performed without errors, the inherent errors in the operands will lead to significant error in the final result. Given that ATOMU computes condition numbers for atomic operations (e.g., subtraction) to identify significant errors, why can we not instead compute the condition number of the entire function with respect to its inputs? Such an approach would enable us to determine the error amplification factor of input perturbations even under the scenario that all other components (e.g., operations) of the function execution are free of error. A small condition number implies that the function's mathematical formulation does not substantially amplify input errors. Consequently, if significant floating-point errors are observed, they can be mitigated by reformulating the function. Such errors are referred to as original-precision-repairable errors.

\noindent \textbf{OFP-Repair: Identifying original-precision-repairable errors by computing the condition number of the entire program.} Following ACESO~\cite{zou2022oracle}, we first utilize ATOMU~\cite{zou2019detecting} to find substantial floating-point errors. Subsequently, for each significant error, we compute the condition number of the input with respect to the function. If the condition number is below a certain threshold, we consider the error to be original-precision-repairable.

With reference to our example in Section~\ref{Motivation}, the condition number with respect to \(x\) is \(\left|\frac{x(\cos(x + \epsilon) - \cos(x))}{\sin(x + \epsilon) - \sin(x)}\right|\), leading to a value of 3.403417\textbf{917418655}. Since it is difficult to obtain derivatives of some functions, we use the numerical differentiation approach to estimate the derivative: $f'(x)\approx {(f(x+h)-f(x))}/h$, resulting in a value of 3.403417\textbf{5514549854}. As our focus is on the order of magnitude of the condition number rather than precise accuracy, the error introduced by the finite difference method is negligible for our purpose. It is noted that the condition number is \(9.0132 \times 10^{9}\) if we take the \(\sin(x + \epsilon)\) as the variable. Therefore, the small condition number we compute indicates that it is possible to mitigate the detected significant error using the original input \(x\), without increasing the floating-point precision.

\noindent \textbf{Our method enables developers to assess whether the SciPy issue \#6368 (cf. subsection~\ref{why_our_method}) can be repaired by using original precision floating-point numbers.} By using OFP-Repair, the condition numbers respect to the two operands are about \(2 \times 10^{30}\) and \(2 \times 10^{5}\), respectively. Therefore, we can confirm that this bug necessitates the implementation of high-precision floating-point arithmetic for repair.

Since OFP-Repair does not utilize high-precision programs and identifies the original-precision-repairable errors successfully, we can overcome limitation 1 and 3.

\subsection{(Mitigating limitation 1 and 2) Step 2: Repair}
\label{step2}

This subsection first introduces the design rationale, followed by the procedure used for repair.

\noindent \textbf{Design rationale: How can the original-precision-repairable program be transformed into one that eliminates the primitive cancellation?} We have shown that significant floating-point errors stem from cancellations (cf. Subsection~\ref{foundations_of}) and our goal of repair is to avoid these cancellations. We also demonstrate that the floating-point program can be transformed into polynomials with respect to the inputs. Therefore, we aim to transform original-precision-repairable programs into polynomial representations that are free of cancellations.

\noindent \textbf{OFP-Repair: Repairing original-precision-repairable errors by using series expansions.} We utilize Taylor series to transform the original programs into polynomials. This choice is motivated by three primary reasons. First, the erroneous inputs lie within a small neighborhood~\cite{zou2019detecting}, where the Taylor series is used to estimate functions near one point. Second, the Taylor series converge rapidly~\cite{stewart2012calculus}, making it computationally efficient. Finally, the terms in the Taylor expansion differ significantly in magnitude, which prevents cancellation from occurring (if the program is original-precision-repairable). Consequently, the primitive function is transformed into: \(\frac{\cos(x)}{1!} (\epsilon) + \frac{-\sin(x)}{2!} (\epsilon)^2 + \frac{-\cos(x)}{3!} (\epsilon)^3 + \frac{\sin(x)}{4!} (\epsilon)^4 + \frac{\cos(x)}{5!} (\epsilon)^5 \cdots\), which avoids the primitive cancellation and mitigates the error. 

Since our repair process does not rely on high-precision programs and can be generalized to many cases based on our theoretical analysis and assumption, OFP-Repair can mitigate limitation 1 and 2. In the following section, we will evaluate the generality.

\noindent \textbf{The limitation of our method:} The Taylor expansion requires the function to be convergent at the expansion point. Consequently, our method becomes inapplicable when the function diverges at the expansion point. For instance, as demonstrated in SciPy issue \#3545, the term \(norm.ppf(1 - q / 2)\) diverges as \(q\) approaches zero.
\section{Evaluation}
\label{eva}

In Section~\ref{OFP-Repair}, we demonstrate that OFP-Repair can address the three limitations of existing methods to repair original-precision-repairable errors: 1) Either the detection or the repair part requires high-precision programs; 2) Can only address a fraction of such errors; 3) Inability to identify such errors. In this section, we perform empirical evaluations to assess the effectiveness of our method in detecting and repairing the original-precision-repairable errors.

Our evaluation begins with a detailed presentation of the experimental setup, covering the datasets, evaluation metric, and our experimental environment. Subsequently, we present the experimental results to illustrate that our method can successfully detect and repair the original-precision-repairable errors. Since ACESO~\cite{zou2022oracle} is currently the only existing tool that does not require high-precision programs for repair, we select it as the baseline approach in our work.


\subsection{Experiment setup}

\subsubsection{Dataset}

Our evaluation is conducted using datasets from ACESO~\cite{zou2022oracle}, which comprise 32 benchmark functions. Among these, 15 functions have been employed in prior floating-point error analysis and repair studies~\cite{panchekha2015automatically, solovyev2018rigorous, wang2019global}, where they were successfully corrected using original-precision floating-point numbers. The remaining 17 functions are variants of original-precision-repairable functions of the literature~\cite{hamming2012numerical} that incorporate library calls from GSL and ALGLIB, making it challenging to construct high-precision versions for them. Consequently, oracle-based repair methods~\cite{yi2019efficient} are difficult to deploy on these functions. Additionally, due to the presence of external function calls, rewriting-based approaches~\cite{panchekha2015automatically} are also less applicable. Overall, all 32 functions can be repaired using original-precision floating-point numbers. By analyzing the bugs and patches collected by Franco et al.~\cite{di2017comprehensive}, we identify five original-precision-repairable errors. We use these cases to assess the performance of our approach on real-world bugs. 

Although our method does not require high-precision oracles during either error detection or repair, we leverage the high-precision programs of ACESO as ground truth to validate the error reduction capability of our patches. For real-world cases, we adopt the patches provided by the developers.

\subsubsection{Evaluation metric}
We utilize relative error to measure floating-point errors: \(\left| \frac{Result_{\text{approximate}} - Result_{\text{true}}}{Result_{\text{true}}} \right|\), where \(Result_{\text{approximate}}\) represents for the results of patches and \(Result_{\text{true}}\) denotes the ground truth from high-precision programs.

\subsubsection{Software and hardware environment}
Our experimental environment is a Docker container running Ubuntu 24.04 on a system with a 13th Gen Intel(R) Core(TM) i9-13900K CPU and 128GB RAM. 

\subsection{Evaluation results}
This subsection presents the experimental results of evaluating our proposed method, OFP-Repair, structured around three key research questions (RQs). For each RQ, we first outline its motivation and the corresponding evaluation methodology, then provide a comprehensive analysis of the results.

\subsection*{RQ1 (addressing limitation 1 and 3): Can OFP-Repair identify original-precision-repairable errors?}
\label{rques1}

\noindent\textit{\textbf{Motivation.}} OFP-Repair employs numerical differentiation to compute program derivatives with respect to inputs, which are then used to compute condition numbers. When an input produces significant floating-point errors but exhibits a small corresponding condition number, we classify such errors as original-precision-repairable. To evaluate our method's capability in detecting these repairable errors, we implement OFP-Repair on all 32 functions from ACESO. It is noted that these 32 cases are original-precision-repairable. Since numerical differentiation may yield derivatives with significant errors, we conduct experiments to demonstrate that these inaccuracies do not compromise the method's effectiveness in identifying original-precision-repairable errors.

\noindent\textit{\textbf{Approach.}} The datasets of ACESO~\cite{zou2022oracle} comprise 32 functions along with narrow input ranges that induce significant errors. The midpoint of each range yields the maximum error, indicating that all points within these ranges result in extremely large condition numbers for certain operations in the program. For instance, in Section~\ref{overcoming_limitation_1and3}, our illustrative example demonstrates a corresponding condition number of \(9.0132 \times 10^{9}\). In ATOMU~\cite{zou2019detecting}, a condition number of \(1 \times 10^{5}\) is considered exceptionally large. To approach the midpoint and trigger a significant error of the function result, we compute the condition number for points at a distance with a radius of \(1 \times 10^{-5}\) and examine whether its corresponding condition number is significantly smaller than \(1 \times 10^{5}\). Subsequently, we compare the analytical solution derived using derivative rules to assess whether the error in our numerical derivative computation affects the detection performance of our method.

\noindent\textit{\textbf{Results. }}\textbf{OFP-Repair effectively identifies the original-precision-repairable errors.} Among these 32 functions, the maximum, minimum, and average values of the computed condition numbers are 1.47, 0, and 0.31, respectively. All computed values are significantly smaller than \(1 \times 10^{5}\). Furthermore, since our sampling points are positioned extremely close to range midpoints, the condition numbers of operations susceptible to cancellation should theoretically be very large (exceeding \(1 \times 10^{5}\)). Therefore, the computed condition numbers of input-to-function mappings are sufficiently small to reliably identify these original-precision-repairable errors.

\textbf{The precision of numerical derivatives has negligible impact on OFP-Repair's detection effectiveness.} Table~\ref{numerical_derivative} demonstrates the results of our numerical derivatives and the corresponding relative errors. The largest relative error occurs with bj\_tan, as its analytical derivative equals 0---theoretically suggesting an extremely small condition number at this point. Excluding this case, the observed relative errors range from 0 to 0.746. These errors do not affect our condition number estimation, as only order-of-magnitude precision is required rather than exact values. Consequently, numerical derivative errors have no material impact on our method's identification accuracy.

\begin{table*}[]
\caption{Numerical derivatives and relative errors compared to analytical solutions. (Note: We set the relative error of bj\_tan to 1 because the analytical value is 0. For cos\_x2, the relative error is 0 because both analytical and numerical derivatives are 0.} 
\vspace{-0.1cm}
\footnotesize
\label{numerical_derivative}
\begin{tabular}{lrrlrrlrrlrr}
\cmidrule(lr){1-3} \cmidrule(lr){4-6} \cmidrule(lr){7-9} \cmidrule(lr){10-12} 
\multicolumn{1}{c}{\textbf{Function}} & \multicolumn{1}{c}{\textbf{Numerical}}  & \multicolumn{1}{c}{\textbf{Relative}} & \multicolumn{1}{c}{\textbf{Function}} & \multicolumn{1}{c}{\textbf{Numerical}}  & \multicolumn{1}{c}{\textbf{Relative}} & \multicolumn{1}{c}{\textbf{Function}} & \multicolumn{1}{c}{\textbf{Numerical}}  & \multicolumn{1}{c}{\textbf{Relative}} & \multicolumn{1}{c}{\textbf{Function}} & \multicolumn{1}{c}{\textbf{Numerical}}  & \multicolumn{1}{c}{\textbf{Relative}} \\
\multicolumn{1}{c}{\textbf{Name}}     & \multicolumn{1}{c}{\textbf{Derivative}} & \multicolumn{1}{c}{\textbf{Error}}    & \multicolumn{1}{c}{\textbf{Name}}     & \multicolumn{1}{c}{\textbf{Derivative}} & \multicolumn{1}{c}{\textbf{Error}}    & \multicolumn{1}{c}{\textbf{Name}}     & \multicolumn{1}{c}{\textbf{Derivative}} & \multicolumn{1}{c}{\textbf{Error}}    & \multicolumn{1}{c}{\textbf{Name}}     & \multicolumn{1}{c}{\textbf{Derivative}} & \multicolumn{1}{c}{\textbf{Error}}    \\
\cmidrule(lr){1-3} \cmidrule(lr){4-6} \cmidrule(lr){7-9} \cmidrule(lr){10-12} 
exp\_BI  & 1.25E-01   & 8.88E-06          & Si\_tan  & -8.47E-02  & 5.00E-01 & fc\_bj   & 5.00E-01   & 1.79E-06          & exp\_1    & 1.00E+00   & 5.00E-06 \\
bJ\_sin  & 2.50E-01   & 2.39E-06          & by\_psi  & 2.62E+00   & 1.66E-01 & cos\_x2  & 0.00E+00   & \textbf{0.00E+00} & x\_tan    & 5.95E-01   & 4.34E-06 \\
di\_tan  & 2.34E-01   & 2.09E-02          & fdm\_log & 2.50E-01   & 1.59E-06 & exp\_2   & 1.00E+00   & 1.67E-06          & log\_log  & -1.00E+00  & 5.56E-06 \\
log\_erf & -1.20E+00  & 2.94E-06          & eQ\_sqrt & 1.25E-01   & 1.32E-06 & cos\_sin & 5.00E-01   & 0.00E+00          & log\_x    & -2.00E+00  & 1.31E-10 \\
acos\_fd & 2.50E-01   & 2.81E-05          & W\_var   & -4.60E-02  & 1.27E-05 & sin\_sin & -1.55E-11  & 4.76E-01          & sqrt\_exp & 3.54E-01   & 3.81E-06 \\
ei       & 8.86E-01   & 9.48E-06          & W\_log   & -3.07E-02  & 1.87E-05 & tan\_tan & 3.10E-11   & 4.76E-01          & sin\_tan  & 1.91E-01   & 7.46E-01 \\
Q1\_W    & 2.00E+00   & 3.20E-06          & pow\_df  & -1.36E+00  & 5.03E-06 & cos\_cos & -1.00E-06  & 1.10E-05          & exp\_x    & 5.00E-01   & 2.22E-06 \\
bj\_tan  & -1.67E-06  & \textbf{1.00E+00} & chi\_ci  & -8.88E-01  & 5.00E-01 & exp\_exp & 2.00E+00   & 0.00E+00          & x\_x2     & -1.00E+00  & 1.00E-05 \\
\cmidrule(lr){1-3} \cmidrule(lr){4-6} \cmidrule(lr){7-9} \cmidrule(lr){10-12} 
\end{tabular}
\vspace{-0.2cm}
\end{table*}

\findingboxx{Our method effectively detects original-precision-repairable errors without high-precision programs. This capability directly addresses limitations 1 and 3 of existing tools (as outlined in Section~\ref{Motivation}).}

\subsection*{RQ2 (addressing limitation 1 and 2): Can OFP-Repair repair original-precision-repairable errors?}
\label{rques2}

\noindent\textit{\textbf{Motivation.}} OFP-Repair utilizes Taylor series to convert the programs into polynomials to obtain patches. To validate the repair effectiveness of our method, we conduct comparative experiments against ACESO, demonstrating that OFP-Repair not only enhances the original program's accuracy but also outperforms ACESO in error correction. 

\noindent\textit{\textbf{Approach.}} For a rigorous comparative evaluation between the patches generated by our method and those produced by ACESO, we adopt identical experimental procedures to those used in ACESO's implementation. The ACESO dataset comprises 32 functions along with input ranges that induce large errors in these functions. The radius of these ranges is 0.01, with the midpoint corresponding to the point of maximum error. Their evaluation approach involves sampling points within these ranges and computing their corresponding errors, ultimately selecting the maximum value as the final result (where lower maximum values indicate better repair performance). Two distinct sampling strategies were employed: (1) focused sampling near range midpoints which can lead to extremely substantial errors (`on Decayed Area'), and (2) uniform sampling across the entire range (`on Stable Area'). In our experiments, at most ten terms of the Taylor series are retained, because the subsequent term can be canceled out. The maximum interval is 0.01, and the 10th power of 0.01 is \(1\times10^{-20}\). In double-precision floating-point arithmetic, adding \(1\times10^{-20}\) to 0.01 still results in 0.01 due to finite precision.

\begin{figure*}[htbp]
    \centering
    \includegraphics[width=0.98\textwidth]{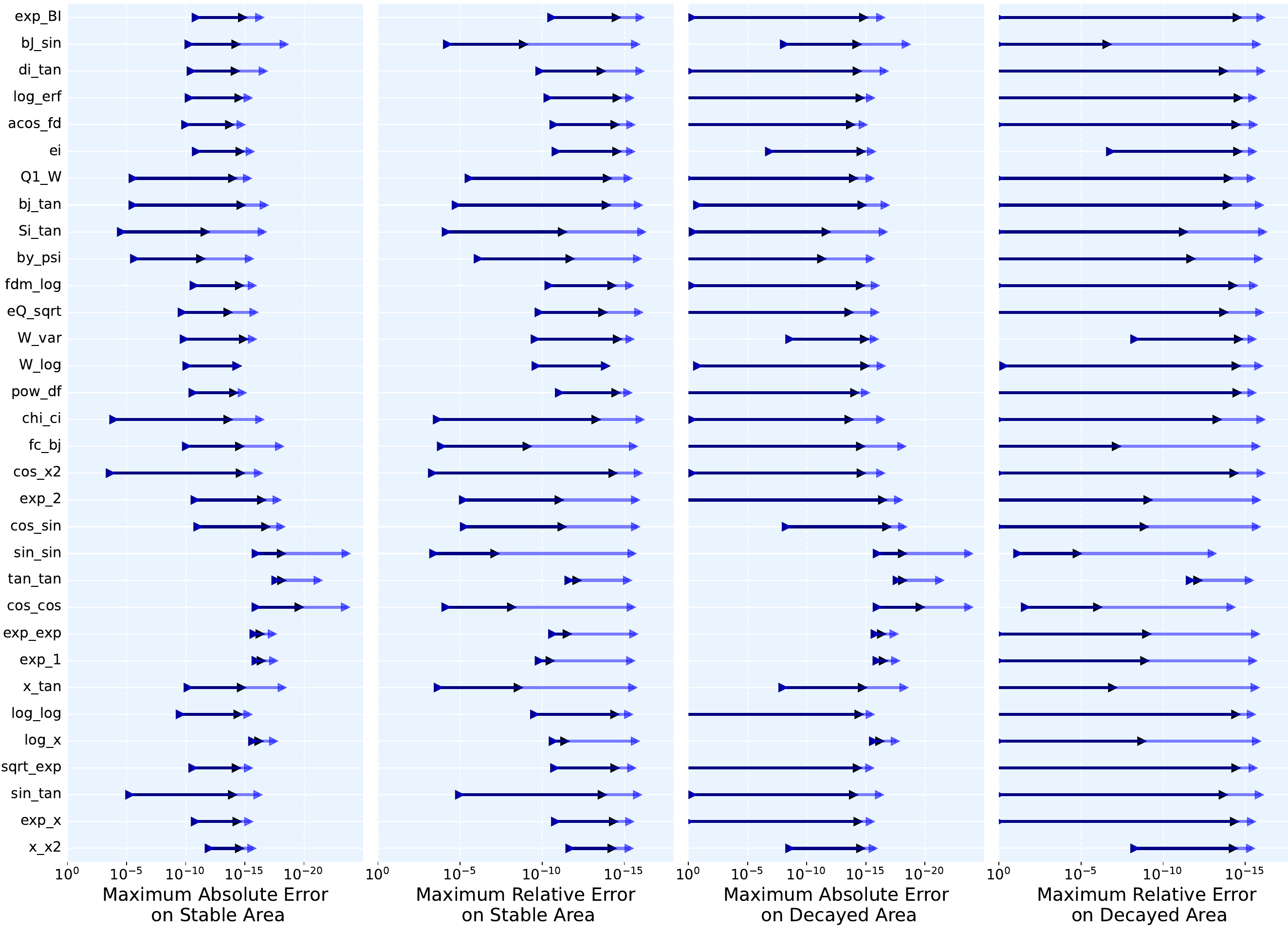}
     \vspace{-0.2cm}
    \caption{Accuracy improvements by ACESO and our method. Results are presented in terms of both absolute and relative errors, covering both the stable and decayed regions of the input ranges. The black arrowed lines originate from the maximum error of the original programs under our test inputs, with arrowheads indicating the maximum error after ACESO's fixing. Similarly, the blue arrowed lines depict the original maximum errors pointing to the errors improved by our proposed method.}
    \label{compare}
\end{figure*}

\noindent\textit{\textbf{Results. }}\textbf{OFP-Repair successfully repair the original-precision-repairable errors.} Figure~\ref{compare} illustrates our method consistently outperforms ACESO across all cases. Specifically, for the maximum absolute error on the stable area, maximum relative error on the stable area, maximum absolute error on the decayed area, and maximum relative error on the decayed area, the average values obtained by our method across all functions are \(4.11\times 10^{-16}\), \(7.47\times 10^{-16}\), \(2.13\times 10^{-16}\), and \(3.73\times 10^{-15}\), repectively. In comparison, ACESO yields \(2.45 \times 10^{-13}\), \(2.74 \times 10^{-9}\), \(2.45\times 10^{-13}\), and \(5.74\times 10^{-07}\) for the same metrics. Our method shows precision improvements over ACESO by three, seven, three, and eight orders of magnitude, respectively.

\textbf{OFP-Repair exhibits robust error mitigation, maintaining consistent effectiveness regardless of the function type.} Since the magnitude of absolute errors scales with the result values, we focus our analysis on relative errors. Across both stable and decayed regions, our method reduces the relative errors to approximately \(10^{-16}\) for the majority of functions (with the exception of \(W\_log\) and \(sin\_sin\)). In contrast, ACESO exhibits unstable performance, with relative errors exceeding \(10^{-10}\) for certain functions. Stable repair is critical in practice, as the absence of ground-truth oracles makes it impossible to empirically validate patching effectiveness. Unstable methods risk leading developers to adopt low-quality patches, which may consequently induce severe operational faults.

\findingboxx{Our method repairs floating-point errors while maintaining original-precision across all target functions, eliminating the need for high-precision implementations. This advancement specifically overcomes limitations 1 and 2 (cf. Section~\ref{Motivation}) inherent in current fixing approaches.}

\subsection*{RQ3 Can OFP-Repair be applied to errors in real-world projects?}
\label{rques3}

\noindent\textit{\textbf{Motivation.}} In RQ1 and RQ2, we demonstrate that our method performs well in both detection and repair on the ACESO dataset. Although this dataset has been widely used in prior research~\cite{panchekha2015automatically, solovyev2018rigorous, wang2019global}, the errors it contains do not originate from real-world software projects. To further validate the effectiveness of our approach, we apply our method to five original-precision-repairable errors collected by Franco et al.~\cite{di2017comprehensive} from real-world programs. 

\noindent\textit{\textbf{Approach.}} We follow the settings and procedures outlined in RQ1 and RQ2. Among these five bugs, SciPy issue \#4034, \#3547, and \#3545 correspond to 2, 1, and 2 cases, respectively. All of them exhibit significant errors when the input values approach zero. Therefore, we select zero as the midpoint and construct ranges with a radius of 0.01. We first compute the condition number for each function when the input is \(1 \times 10^{-5}\). It should be noted that for the first case of SciPy issue \#3545, the function value equals zero when the input is \(1 \times 10^{-5}\). Consequently, we set the input to 0.1 instead. Subsequently, we employ our method to repair these five errors.

\noindent\textit{\textbf{Results. }}\textbf{OFP-Repair successfully identify all the five cases.} The five computed condition numbers are 3.00, 0.41, 0.03, 5.43, and 0.03, all of which are significantly below \(1 \times 10^{5}\). Thus, our method successfully identifies them as original-precision-repairable. Since developers may sometimes be uncertain whether a particular error necessitates high-precision fixes(cf. Section~\ref{Motivation}), OFP-Repair can provide them with a reliable determination.

\begin{figure*}[htbp]
    \centering
    \includegraphics[width=0.98\textwidth]{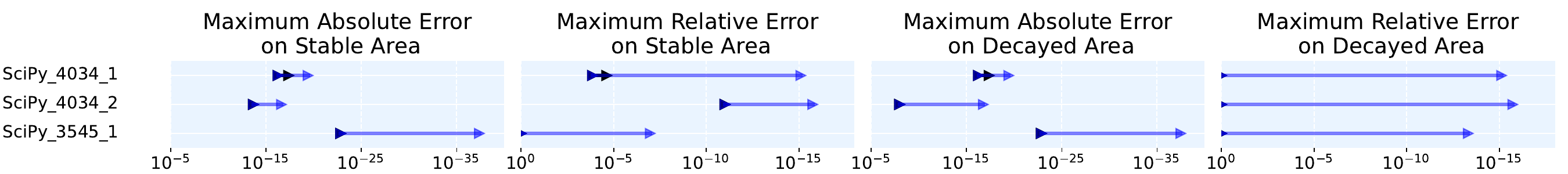}
     \vspace{-0.2cm}
    \caption{Performance comparison between ACESO and our method on three real-world cases. }
    \vspace{-0.2cm}
    \label{compare_rq3}
\end{figure*}

\textbf{OFP-Repair can successfully repair three of these bugs while ACESO can only fix one case.} As shown in Figure~\ref{compare_rq3}, ACESO is only effective for the three subfigures corresponding to the first function. Moreover, the patches generated by our method achieve higher numerical accuracy. These results demonstrate the superior reliability of our method for repairing such floating-point errors in numerical software.

\findingboxx{Our method successfully detects all five original-precision-repairable errors. While the limitation of our Taylor expansion-based approach allows us to repair three functions, ACESO is only able to repair one function.}
\section{Case study: OFP-Repair on open bug reports}
\label{case_study}
In Section~\ref{eva}, we show that our method can effectively detect the original-precision-repairable errors and repair them. However, the experiments conducted across our three research questions are all based on validated original-precision-repairable errors, which fail to fully demonstrate the distinctive advantages of our method. Therefore, in this section, we deploy our method to address bugs documented in an open report dating back over a decade~\footnote{\protect\url{http://savannah.gnu.org/bugs/?43259}}. This bug report contains 15 floating-point errors of GSL functions, and our method successfully identifies and provides precision improvements by using original-precision arithmetic for five of these bugs:

\noindent\textit{\textbf{1) Case 1: Fully resolved: }} The first example illustrates cases where our method can both identify the original-precision-repairable bugs and enhance precision using original-precision floating-point numbers to achieve a relative error below \(1\times 10^{-16}\). Three of the five bugs we can repair fall into this category.

\(gsl\_sf\_hyperg\_0F1(c, x)\) results in a relative error of 1.15 \(\times 10^{198}\) when the two operands are \(3.39\times 10^{-215}\) and \(3.95\times 10^{-242}\). However, the condition numbers associated with the two parameters are \(3.39 \times 10^{-210}\) and \(1.01 \times 10^{-225}\), respectively. Therefore, we identify the bug as the original-precision-repairable case. When \(c\) and \(x\) approach zero, the first two terms in the Taylor expansion of the function are \(1+x/a\). We use this as a patch, and the resulting relative error is only \(1.17 \times 10^{-27}\).

\noindent\textit{\textbf{2) Case 2: Partially addressed: }} This category contains two bugs, indicating that our method can identify cases where the accuracy can be improved using original-precision floating-point arithmetic. However, the relative error can only be reduced to just under \(1 \times 10^{-10}\). For the first bug, \(gsl\_sf\_gamma\_inc\_Q(a, x)\) leads to a significant error if the operands are \(4.57 \times 10^{-13}\) and \(5.13 \times 10^{-61}\). Although we can confirm that this bug can be fixed using original-precision floating-point computation, the function is undefined at zero, rendering our Taylor expansion repair inapplicable (cf. Section~\ref{step2}). However, a part of the function can be well-approximated using a Taylor expansion. We apply our repair strategy to this part, reducing the relative error from \(1.60 \times 10^{-6}\) to \(1.57 \times 10^{-7}\). Concerning the second bug, \(gsl\_sf\_ellint\_P(\phi, k, n)\) produces a relative error as large as \(9.52 \times 10^{6}\). Although the largest of the three condition numbers is \(3.28\times 10^{5}\), we believe that such a condition number should not theoretically result in an error of this magnitude. Upon inspecting the source code, we identify the root cause: the original implementation computes \(\sin(\phi)^3\) directly, but due to \(\phi\) being extremely small, the result underflows to zero. To improve accuracy, we modified the computation by separating the expression---multiplying \(\sin(\phi)\) with a large variable first before cubing. This change successfully reduces the relative error to \(1.91 \times 10^{-9}\).

\section{Threats to Validity}
\label{threat}
In this section, we examine the potential threats to the validity of our study.

\noindent \textbf{External validity.} One potential threat is that the effectiveness of our method may be limited to specific types of data. To address this concern, we conduct evaluations on three distinct datasets: a widely adopted academic benchmark (RQ1 and RQ2), real-world bugs from practical projects (RQ3), and floating-point errors identified in open bug reports (cf. Section~\ref{case_study}). Experimental results demonstrate that our method performs well across all datasets.

\noindent \textbf{Internal validity.} In our experiment, we set the upper limit of Taylor expansion terms to 10. A potential threat to internal validity is that if developers prioritize efficiency over precision in certain scenarios and reduce the number of terms, the accuracy of our generated patches may decrease. However, our configuration is justified based on the characteristics of our dataset. Specifically, our data requires that the input range of the patch spans a maximum distance of 0.01 from the point of highest error---a relatively large interval that necessitates a higher number of terms to maintain precision. In cases where the distance is significantly smaller---for instance, in \(gsl\_sf\_sin\_e\) (cf. subsection~\ref{taylor_series}), where the absolute value of the input is constrained to 1.22e-04---only two expansion terms are sufficient to achieve the desired accuracy.


\noindent \textbf{Construct validity.} In Section~\ref{eva}, our experiments only demonstrate that our detection achieves a high true positive rate; they do not provide evidence that it yields a high true negative rate. Obtaining ground truth for bugs that can only be fixed using high-precision arithmetic is challenging, as developers typically do not provide explicit clarification. Such bugs are often left open indefinitely or closed without resolution. However, in Section~\ref{case_study}, we apply our method to bugs reported in open bug reports. We successfully identify and repair five bugs whose accuracy can be improved using original-precision floating-point arithmetic. This indicates that our method demonstrates a high true negative rate---at least higher than manual judgment. After all, if developers had recognized that these bugs could be fixed with original-precision arithmetic, they would likely have done so rather than leaving them unresolved for over one decade.
\section{Related Work}
\label{related-work}
This section outlines related research directions in error detection and repair.

\noindent \textbf{\textit{Detecting errors of floating-point programs.}} Valgrind~\cite{nethercote2007valgrind} is a dynamic binary instrumentation framework designed to support the construction of heavyweight dynamic analysis tools. Several techniques have been developed on top of Valgrind, including FPDebug~\cite{benz2012dynamic}, Verrou~\cite{franccois2016verrou} and Herbgrind~\cite{sanchez2018finding}. FPDebug employs randomized testing and shadow execution with higher-precision variables to detect catastrophic cancellations and rounding errors. Verrou leverages a Monte Carlo–based approach to identify errors. Herbgrind is designed to trace the root causes of numerical errors in large-scale programs by performing random sampling. Recently, several approaches have been introduced to identify inputs that cause significant floating-point errors~\cite{chiang2014efficient, zou2015genetic, zou2019detecting, yi2019efficient, guo2020efficient, wang2022detecting, zhang2023eiffel, zhang2024hierarchical, yi2024fpcc}. The motivation behind these methods lies in the observation that only a small subset of inputs typically leads to large errors in a program. Therefore, identifying such inputs is crucial. To this end, various techniques have been explored, including genetic algorithms, condition number analysis, and symbolic execution.

\noindent \textbf{\textit{Repairing floating-point errors.}} Several studies have aimed to enhance the precision of floating-point computations by applying mathematically equivalent code transformations~\cite{panchekha2015automatically, damouche2017salsa, xu2024arfa, wang2019global}. However, these methods rely on high-precision computations to obtain an oracle that guides the transformations. Moreover, they are not capable of identifying all bugs that can be repaired using original-precision floating-point arithmetic. AutoRNP~\cite{yi2019efficient} uses high-precision programs to obtain oracles and employs linear approximation to estimate the correct function. To the best of our knowledge, ACESO~\cite{zou2022oracle} is the most closely related approach to ours. It does not rely on high-precision computation in either the detection or repair phase. However, ACESO performs poorly in many cases and cannot identify original-precision-repairable errors.

OFP-Repair aims to detect floating-point errors that can be repaired using original-precision arithmetic and to generate corresponding patches that operate within the same precision.

\section{Conclusion}
\label{conclusion}

Floating-point errors can lead to severe consequences, among which certain types of errors can be repaired using original-precision arithmetic, while others necessitate high-precision solutions. Developers often refrain from addressing the latter due to the substantial resource demands of high-precision implementations. However, existing repair methodologies fail to assist developers in identifying the former category of errors, nor do they provide a unified framework for repair. To address this gap, we propose a novel method, named OFP-Repair. Experimental results demonstrate that our method effectively detects and repairs such errors. When compared to ACESO as a baseline repair algorithm, our patches exhibit significantly superior accuracy. Notably, OFP-Repair successfully resolves five out of 15 bugs in an open bug report.

In future work, we aim to extend the deployment of our method to a broader range of unresolved errors.

\bibliographystyle{ACM-Reference-Format}
\bibliography{main}

\end{document}